\documentclass[a4paper,11pt]{article}
\usepackage{jinstpub}
\usepackage{lineno}
\usepackage{subfig}
\usepackage{wrapfig}
\usepackage{longtable}
\usepackage{booktabs}
\usepackage{caption}
\usepackage{graphics}
\usepackage{graphicx}
\usepackage{mathtools}
\usepackage{chemformula}
\usepackage{circuitikz}
\usepackage{hyperref}
\usepackage[output-decimal-marker={.}]{siunitx}
\proceeding{17$^{\text{th}}$ Topical Seminar on Innovative Particle and Radiation Detectors (IPRD25)\\
15-19 September 2025\\
Siena, Italy}

\title{\boldmath The new truly cylindrical tracker for the ALICE ITS3}

\collaboration[c]{on behalf of the ALICE collaboration}

\author{Stefania Perciballi}
\affiliation{Università di Torino,\\
via Pietro Giuria, 1, Torino, Italy}

\emailAdd{stefania.perciballi@unito.it}

\abstract{The ALICE collaboration is preparing an upgrade of the three innermost layers of the current Inner Tracking System (ITS) during the next LHC long shutdown (LS3). 
The new ITS detector will use wafer-scale (up to \SI{27}{cm} in length) Monolithic Active Pixel Sensors with a \SI{65}{nm} CMOS Image Sensor process, thinned to \SI{50}{\micro m} and bent around the beam pipe. The planned upgrade will allow the use of only two sensors per tracker layer, kept in place by just two mechanical supports at the edges and two thin carbon fibre supports at the sensor border. The substitution of water cooling with air cooling will lead to an expected reduction of the material budget per-layer from $\sim$0.36\% $X_0$ of the current detector to 0.09\% $X_0$. 
The R\&D process also led to the development of a new sensor variant with an additional low dose n-type implant to the previous detector. This improves charge collection speed, confirms a spatial resolution of about \SI{5}{\micro m}, a detection efficiency greater than 99\% and an excellent radiation tolerance.
Large area prototypes proved the possibility to have an active area greater than 90\%, and a fake hit rate lower than \SI{e-6}{hits/pixel/event} without loosing detection efficiency. 
This proceeding will show the above innovations, with particular attention to a small area analogue test structure featuring a front-end which can be monitored via an on-chip Operational Amplifier buffer that preserves the steep signal edge (few hundreds of ps) in order to study the sensor timing performance. The characterization proved a time resolution of \SI{63}{ps} on average and \SI{50}{ps} for signal passing right under the electrode with a detection efficiency above 99\%.
}


\arxivnumber{} % Only if you have one

\begin{document}
\maketitle
\flushbottom

\section{Introduction}

The ALICE experiment at CERN is developing an upgrade to its Inner Tracking System (ITS2), planned for installation during the LHC’s third long shutdown (2026–2030) \cite{ALICE_upgrade_LHC}. The primary objective is to improve the momentum resolution, particularly below \SI{1}{GeV/c}.
\begin{wrapfigure}{r}{0.6\textwidth}
    \centering
    \includegraphics[width=0.52\textwidth]{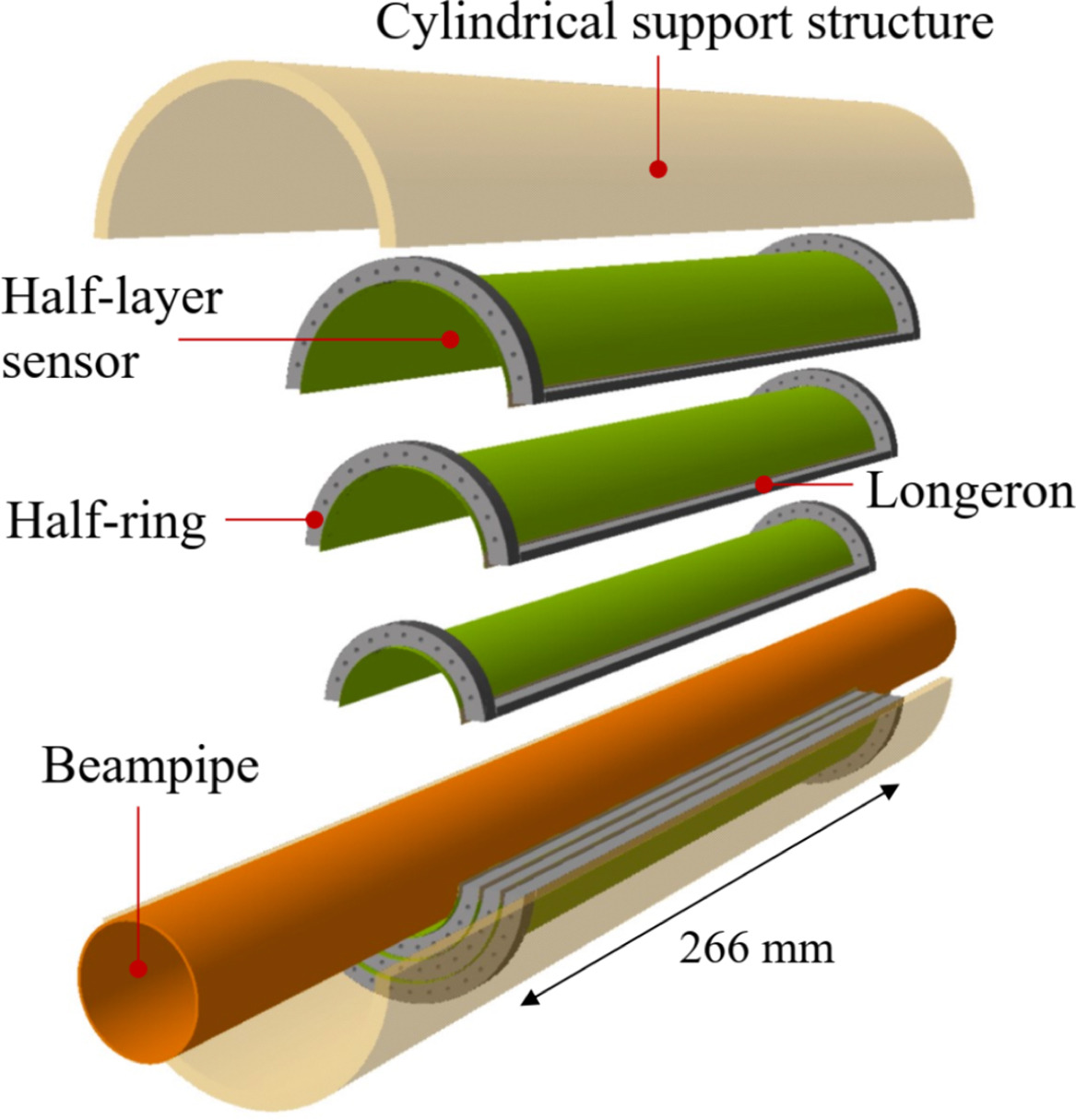}
    \caption{Schematic view of the ITS3 upgrade: it is possible to identify the detector structure composed by six half-cylinder silicon sensors \cite{ITS3_2024_Liu}.}
    \label{fig:ITS3_SchematicView}
\end{wrapfigure}
\noindent The upgrade involves replacing the three innermost layers of the current ITS2, with large area bent silicon sensors, marking the first implementation of a truly cylindrical tracking detector \cite{,ITS3_2024_Liu}.
Each half-layer will be implemented by a single Monolithic Active Pixel Sensor (MAPS) with an area up to $97.8 \times 266 \text{ mm}^2$ manufactured in a commercial \SI{65}{nm} technology developed by Tower Partners Semiconductor Co. (TPSCo) \cite{TPS_Co}.
Moreover, the current water cooling system will be replaced by air cooling while the mechanical stability will be granted by a carbon foam structure placed at the border of each silicon sensor as pictured in figure \ref{fig:ITS3_SchematicView}.
The overall effect will be a material budget reduction from 0.36\% $X_0$ of the ITS2 to 0.09\% $X_0$ per tracking layer.
\begin{figure}[!ht]
    \centering
    \includegraphics[width=\textwidth]{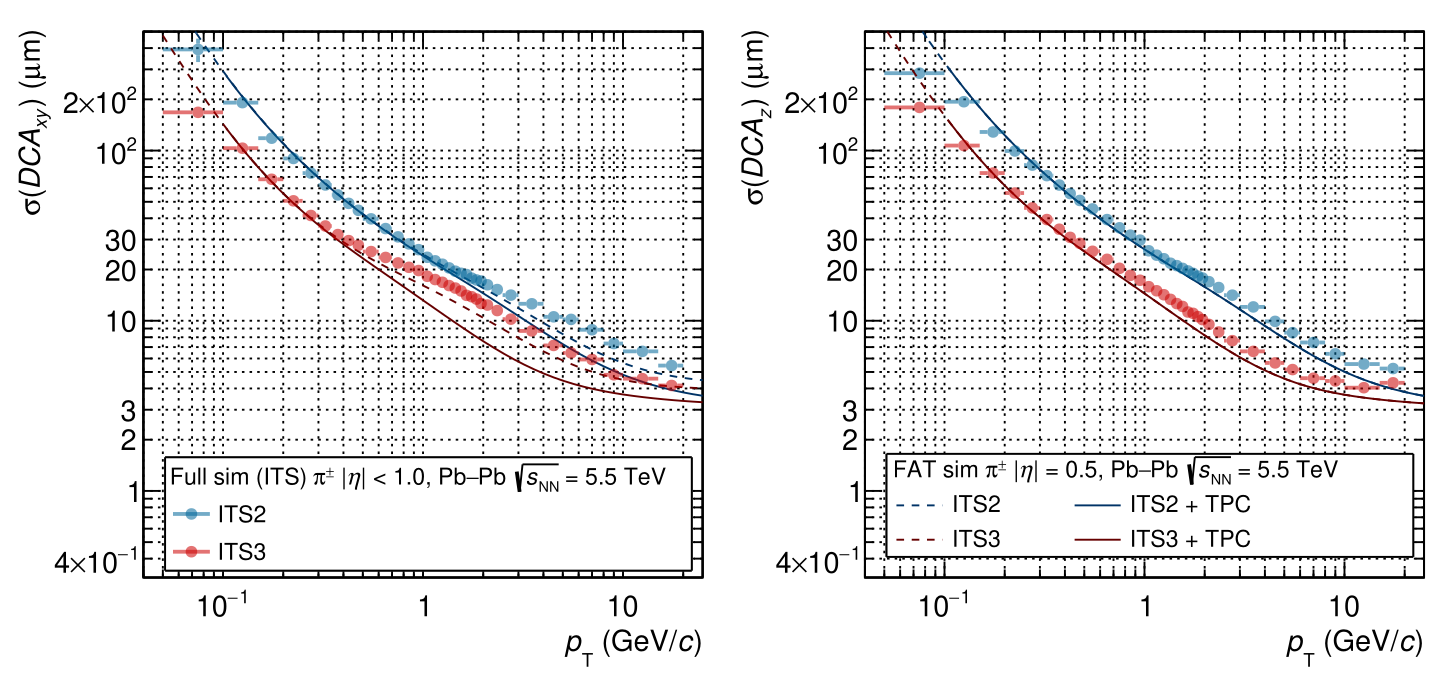}
    \caption{Impact parameter resolution in the xy (left) and z (right) direction of the ITS2, ITS2 + TPC and expected resolution of ITS3, ITS3 + TPC. In both plots, the dots represent the result of a full simulation of the ITS while lines are obtained with a fast simulation (FAT) \cite{ITS3_TDR}.}
    \label{fig:ImpactParameterRes_ITS2-3}
\end{figure}
\vspace{-1pt}
\noindent The proposed upgrade will also reduce the beam pipe radius (from \SI{18.0}{mm} to \SI{16.0}{mm}), consequentially moving the first tracker layer closer to the interaction point (from \SI{22.4}{mm} to \SI{19.0}{mm}). This detector upgrade is referred to as ITS3 \cite{ITS3_TDR}.
From simulations of particle interactions it is possible to evaluate the improvement in the impact parameter resolution obtained by implementing the ITS3 upgrade. Figure \ref{fig:ImpactParameterRes_ITS2-3} reports the simulation results for the ITS2 and ITS3 stand-alone (dotted line) and by adding the TPC contribution in the tracking (full line) on the xy (left) and z (right) component of the impact parameter. Implementing the ITS3 upgrade will improve the resolution by a factor 2 along both directions \cite{ITS3_TDR}.\\
Considerable efforts were directed toward demonstrating the feasibility of this upgrade, as numerous challenges had to be addressed.
Firstly, demonstrate that thin silicon could be bent without breaking and would still be functional, requiring a test on bare thin silicon and then on working silicon sensors, described in section \ref{sec:bending_silicon}.
Secondly, section \ref{sec:sensor} details the process of moving from the well-known \SI{180}{nm} to the \SI{65}{nm} CMOS technology to be able to use \SI{300}{mm} diameter wafers.
Moreover, standard CMOS processes limit the sizes of lithographic masks to the order of $(2\times3\,\mathrm{cm}^2)$. This constraint has recently been overcome using stitching, which replicates mask patterns across a wafer without disrupting electrical continuity, thus allowing the creation of larger sensors \cite{Brevetto_stitching}. Section \ref{sec:stitching_MOSS} is dedicated to the performance of the first stitched sensor prototype.
Lastly, the mechanical question of whether the stand-alone silicon would be robust enough to resist air flow to cool the system and air cooling would be enough to maintain a uniform and stable temperature along the ITS3 is addressed in section \ref{sec:mechanics_cooling} \cite{ITS3_TDR}.
\begin{figure}[!ht]
    \centering
    \includegraphics[width=0.4\textwidth]{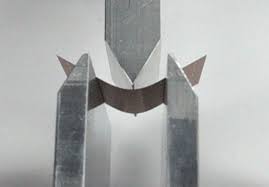}%\quad
    \includegraphics[width=0.58\textwidth]{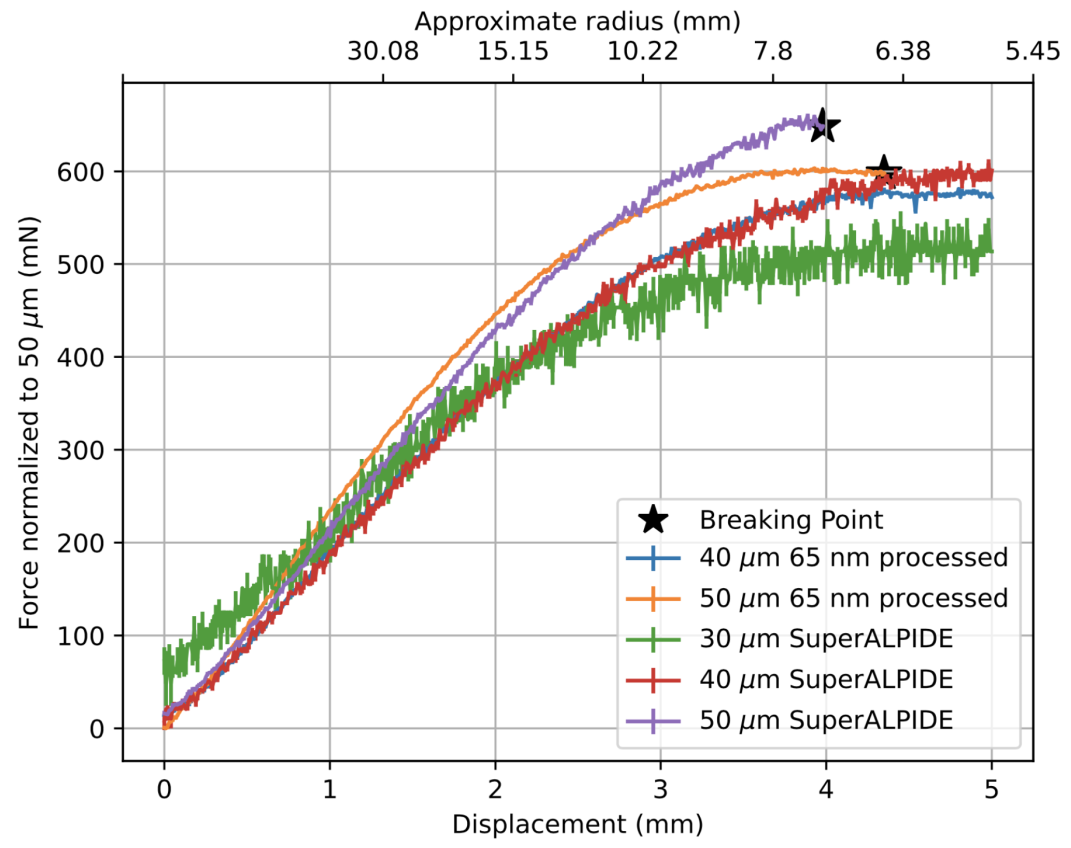}
    \caption{Setup for bending strength measurements (on the left) and bending results on ALPIDE and \SI{65}{nm} sensors (on the right) \cite{ITS3_TDR}.}
    \label{fig:BendSisicon}
\end{figure}
\section{Bending silicon sensors}
\label{sec:bending_silicon}
Firstly, the flexibility of thin silicon was tested by trying to find out which was the maximum force applicable to the sensors before breaking them. Figure \ref{fig:BendSisicon} shows, on the left, the test setup used to bend the chips while, on the right, a graph reports the results obtained with different sensors thicknesses and technologies.
The measurements show that silicon chips with thickness lower than \SI{50}{\micro m} can be bent to the ITS3 target radii (\SI{19}{mm}) without breaking, and that sensors are robust enough to resist mechanical stress \cite{ITS3_2024_Magnus}.
\begin{figure}[!ht]
    \centering
    \includegraphics[width=0.35\textwidth]{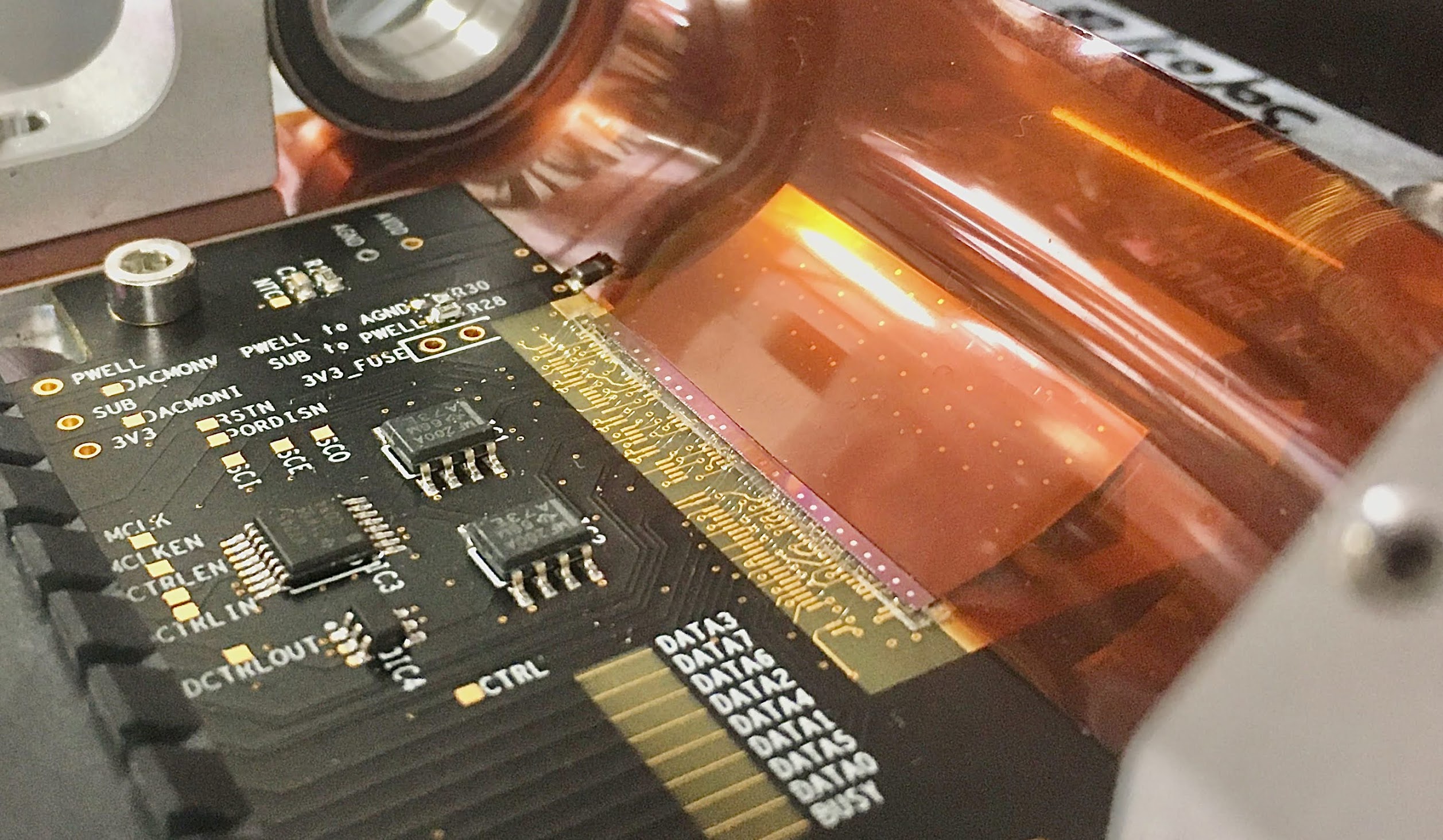}
    \includegraphics[width=0.63\textwidth]{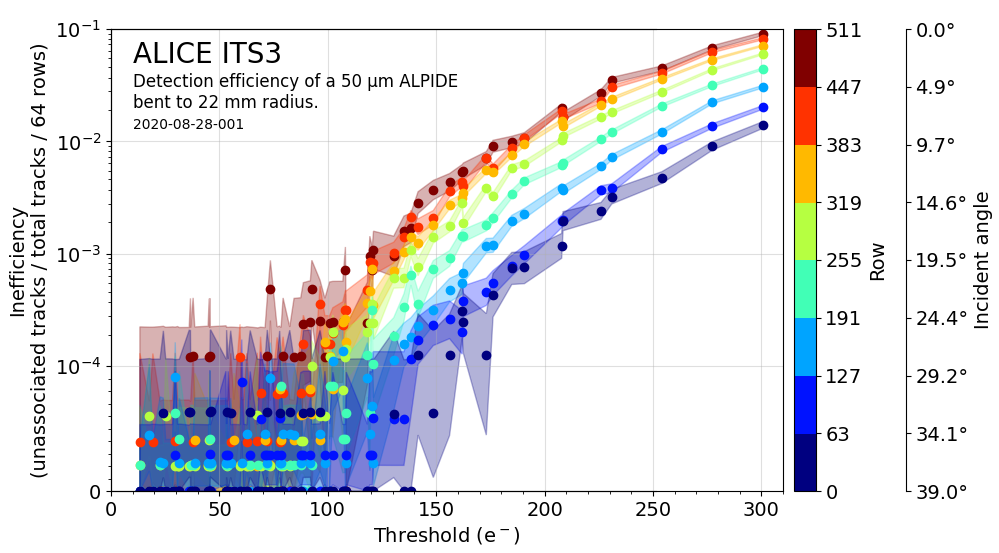}
    \caption{On the left: ALPIDE bent along the short edge and wire bonded to a carrier board. On the right: inefficiency of the ALPIDE chip bent along short axis as a function of threshold for different rows and incident angles \cite{BentALPIDE}.}
    \label{fig:BentALPIDE}
\end{figure}
The next test was to bend an ALPIDE reference sensor along the short edge to a radii of \SI{22}{mm} and test its performance during a campaign with a \SI{5.4}{GeV} electron beam. The results of these measurements are presented in figure \ref{fig:BentALPIDE}. Figure \ref{fig:BentALPIDE} (right) shows that bent ALPIDE becomes less efficient when decreasing the incident angle of the particles, due to the smaller average deposited charge, as expected. However, for thresholds lower than \SI{100}{e^-}, the inefficiency is lower than $10^{-4}$, demonstrating that ALPIDE retains an excellent detection performance after bending \cite{BentALPIDE}.
\begin{wrapfigure}{r}{0.6\textwidth}
    \centering
    \includegraphics[width=0.55\textwidth]{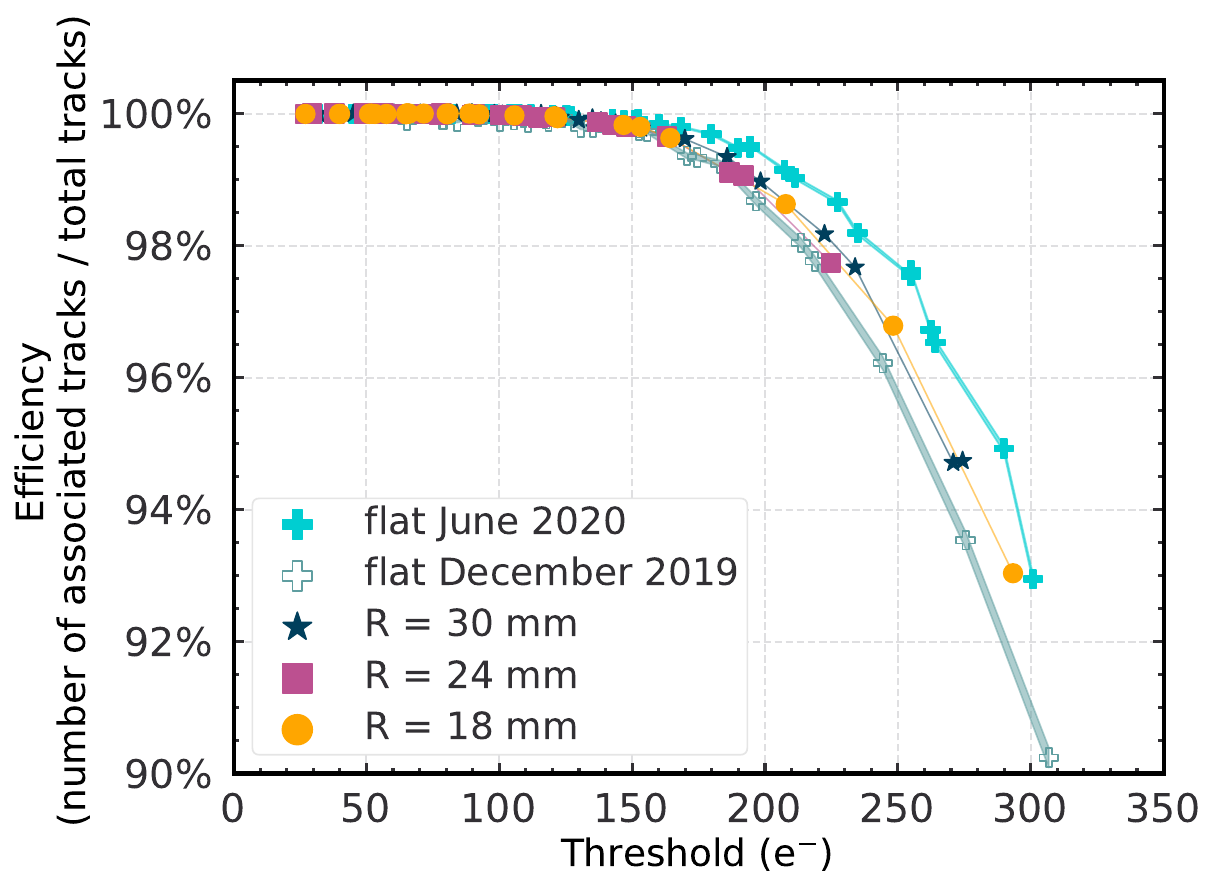}
    \caption{Efficiency of bent ALPIDE sensors as a function of the threshold for the three different bending radii compared to flat sensors \cite{micro_ITS3}.}
    \label{fig:uITS3_eff}
\end{wrapfigure}
Subsequently, the ALPIDE was bent along the long edge, and a mock-up version of the ITS3 was made to prove the tracking performance of bent silicon sensors in the final configuration. A holder was designed to test the small ITS3 version and investigate the tracking performance. A window was left to allow particles to cross the three bent ALPIDE layers.
During 2019-2021, several test beams were made both at DESY and at CERN SPS; the results obtained on those occasions are reported in the plot in figure \ref{fig:uITS3_eff}. Both flat and bent ALPIDEs were tested, and the detection efficiency of ALPIDEs bent to the three ITS3 radii are compatible with the results obtained on flat ALPIDEs \cite{micro_ITS3}. 
\section{Technology validation}
\label{sec:sensor}
\begin{figure}[!ht]
    \centering
    \includegraphics[width=0.3\linewidth]{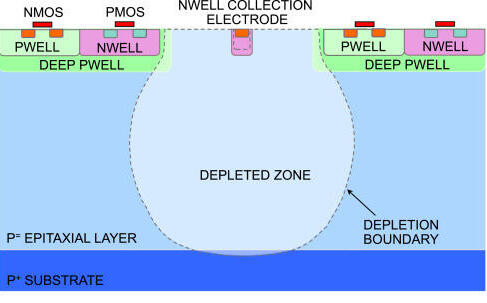} \quad
    \includegraphics[width=0.3\linewidth]{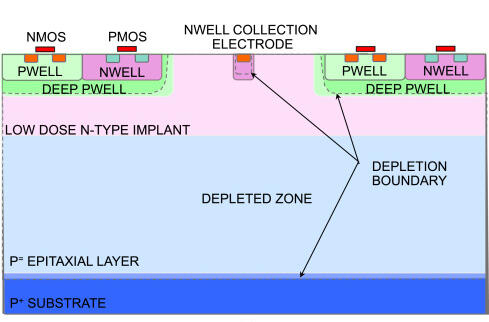} \quad
    \includegraphics[width=0.3\linewidth]{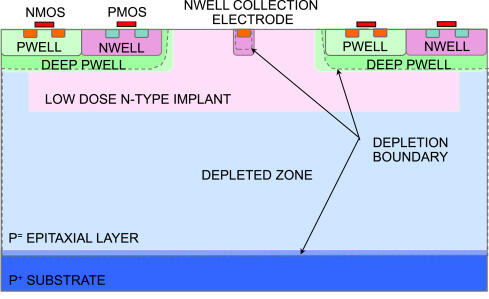}
    \caption{From left to right: standard, modified and modified-with-gap sensor variants \cite{ITS3_TDR}.}
    \label{fig:sensor_variants}
\end{figure}
\begin{wrapfigure}[17]{r}{0.55\textwidth}
    \centering
    \includegraphics[width=0.47\textwidth]{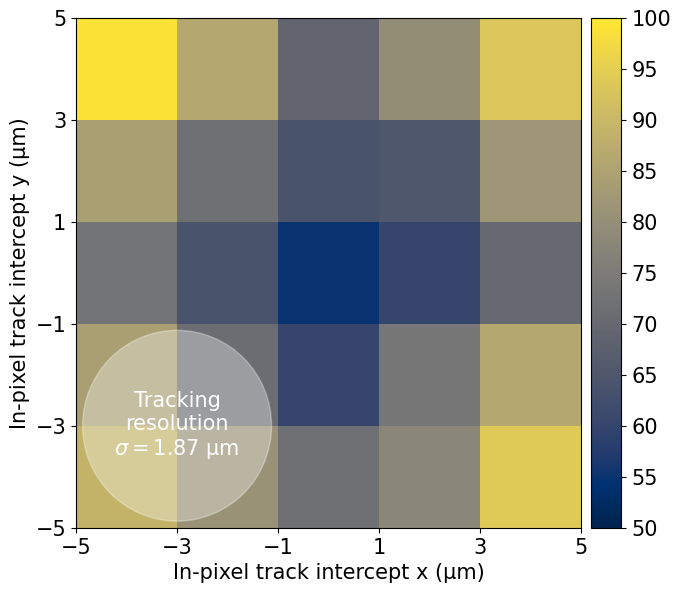}
    \caption{Corrected in-pixel time resolution of the modified-with-gap structure operated at V$_{sub}$=\SI{-4.8}{V} \cite{opamp2025}.}
    \label{fig:TB_inpixel_timeresolution}
\end{wrapfigure}

\noindent The R\&D for the ITS3 sensor originated from the ALPIDE sensor currently used in the ALICE ITS2, but incorporated advancements from silicon detector research. This included a design modification to the standard ALPIDE by introducing a low-dose n-type implant beneath the collection electrode (modified version), aimed at increasing charge collection speed. To further optimize the lateral electric field, a gap was introduced at the pixel borders (modified-with-gap version) \cite{Magdalena_2019}. The three different structures are shown in figure \ref{fig:sensor_variants}. Small area test structures were produced to validate the \SI{65}{nm} CMOS technology, investigate different detector doping profiles, optimize sensor performance, and choose the optimal pixel pitch to achieve a spatial resolution of \SI{5}{\micro m} \cite{Magdalena_2020}. In particular, the first sensor submission was called the Multi Layer Reticle 1 (MLR1), which featured four different types of sensors: 
\begin{itemize}
    \item Circuit Exploratoire 65 (CE65): a $64\times32$ or $48\times32$ pixel matrix with pitches of \SI{15}{\micro m} or \SI{25}{\micro m} and featuring rolling shutter readout.
    \item Digital Pixel Test Structure (DPTS): a $32\times32$ pixel matrix with \SI{15}{\micro m} pitch, optimized to test the full readout chain as it will be designed for ITS3. 
    \item Analogue Pixel Test Structure - Source Follower (APTS-SF): a $4\times4$ pixel structure with \SI{10}{\micro m}, \SI{15}{\micro m}, \SI{20}{\micro m}, and \SI{25}{\micro m} pitches optimized for analogue measurements and variant comparison.
    \item Analogue Pixel Test Structure - Operational Amplifier (APTS-OA): a $4\times4$ pixel matrix with \SI{10}{\micro m} pitch optimized to extract the sensor timing performances independently from the front-end electronics that will be used in the final sensor.
\end{itemize}
\begin{figure}[!ht]
    \centering
    \includegraphics[width=0.95\textwidth]{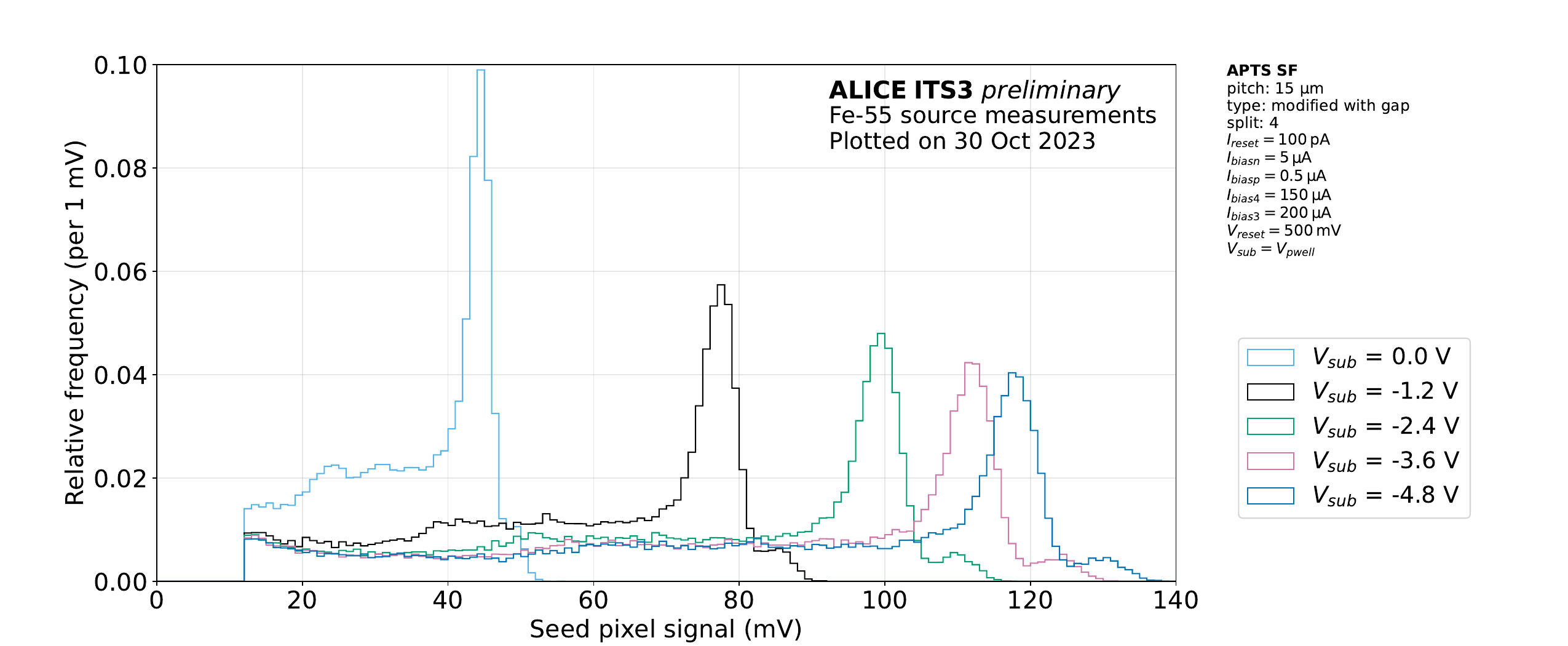}
    \caption{Seed signal distribution comparison among different reverse substrate voltages for a modified-with-gap APTS-SF with \SI{15}{\micro m} pitch \cite{sf2024}.}
    \label{fig:APTS-SF_spectrum}
\end{figure}
\noindent The sensor performance was tested through different measurements both in the lab, at test beam facilities and with different X-ray sources; the measurements were made with multiple combinations of doping concentrations, pixel layouts, and pitches. The APTS-SF was used to study various sensor designs, pixel sizes, and doping concentrations, as detailed in \cite{sf2024}. As an example, figure \ref{fig:APTS-SF_spectrum} shows a $^{55}$Fe spectrum measured with a \SI{15}{\micro m} pitch APTS-SF at different reverse bias voltages, illustrating how signal amplitude increases with substrate voltage due to the expanding depletion region.
\begin{figure}[!ht]
    \centering
    \includegraphics[width=0.95\textwidth]{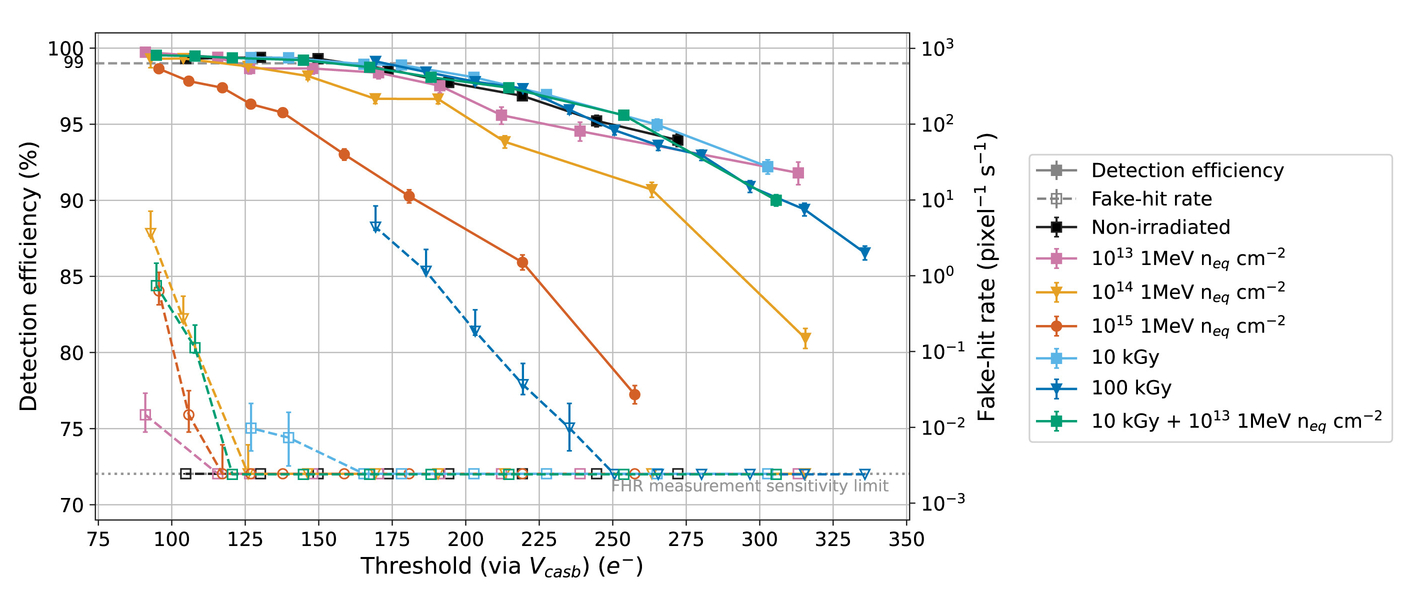}
    \caption{DPTS detection efficiency (filled symbols, solid lines) and fake-hit rate (open symbols, dashed lines) as a function of average threshold, measured with \SI{10}{GeV/c} positive hadrons. \cite{dpts2023}.}
    \label{fig:DPTS_eff_irradiated}
\end{figure}
Meanwhile, the Digital Pixel Test Structure was used to measure the detection efficiency and the fake hit rate of a \SI{65}{nm} CMOS sensor with the associated front-end electronics during test beam measurements.
Figure \ref{fig:DPTS_eff_irradiated} describes the sensor detection efficiency and fake hit rate as a function of the applied threshold. 
The detection efficiency is evaluated as the fraction between tracks associated to an event measured by the DPTS and the total number of tracks passing trough the detector.
It can be observed that non-ionising irradiation leads to a decrease in the detection efficiency, while ionising irradiation leads to an increase in the fake-hit rate \cite{dpts2023}. \\
The detector functionalities for irradiation up to \SI{4e12}{1MeV n_{eq} cm^{-2}} are reached for thresholds greater than \SI{125}{e^-} \cite{ITS3_TDR}. In addition to these structures, the APTS-OA was also included to measure the sensor intrinsic time resolution, independently from the front-end electronics.
\begin{figure}[!ht]
    \centering
    \includegraphics[width=0.8\textwidth]{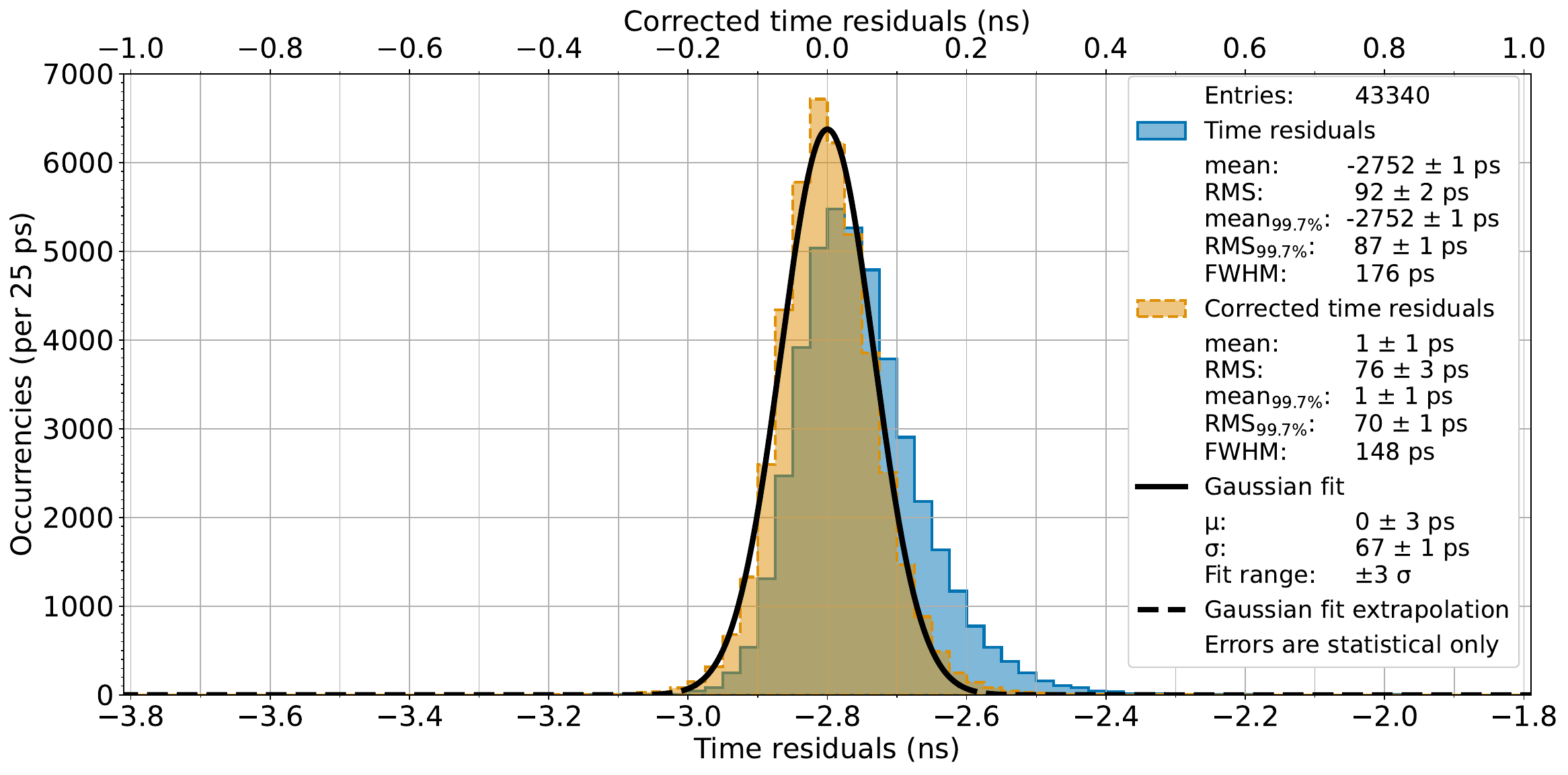}
    \caption{Time residuals measured for the modified-with-gap sensor operated at V$_{sub}$=\SI{-4.8}{V}, before (blue distribution, bottom horizontal axis) and after (orange distribution, top horizontal axis) time walk-like correction. The latter distribution is fitted with a Gaussian function on a range of $\pm 3\sigma$ (black solid line, dashed line for points outside the fit range) \cite{opamp2025}.}
    \label{fig:TB-timeResiduals}
\end{figure}
The APTS-OA structure allowed to measure the time resolution during a test beam performed at SPS at CERN. 
A Low Gain Avalanche Detector (LGAD) produced by Fondazione Bruno Kessler \cite{LGAD_2023} was operated as time reference to evaluate the time resolution of a modified-with-gap APTS-OA with a \SI{10}{\micro m} pitch.\\
The time difference calculated as $\Delta t = t^{\textrm{DUT}}_{10\%} -t^{\textrm{LGAD}}_{40\%} $ is plotted in figure \ref{fig:TB-timeResiduals}.
The absence of a shaper in the front-end results in observable time-walk effects, particularly in the tails of the distribution.
This effect has been corrected as described in \cite{opamp2025}. After time-walk correction, a time resolution of \SI{63}{\pico s} is obtained, calculated as the RMS of the distribution.\\
Figure \ref{fig:TB_inpixel_timeresolution} reports the time resolution as a function of the track crossing position inside the pixel. From the graph it is possible to observe that the best time resolution is obtained for tracks passing right under the pixel electrode and can be as low as \SI{50}{\pico s}, while it increases when the track passes through a region further away from the pixel centre.
\section{The first large area stitched-sensor}
\label{sec:stitching_MOSS}
As part of the R\&D efforts, a large-area prototype detector named the Monolithic Stitched Sensor (MOSS) was developed in 2023. To produce elements exceeding the reticle size, a stitching process is used.
\begin{figure}[b]%[!ht]
    \centering
    \includegraphics[width=0.73\textwidth]{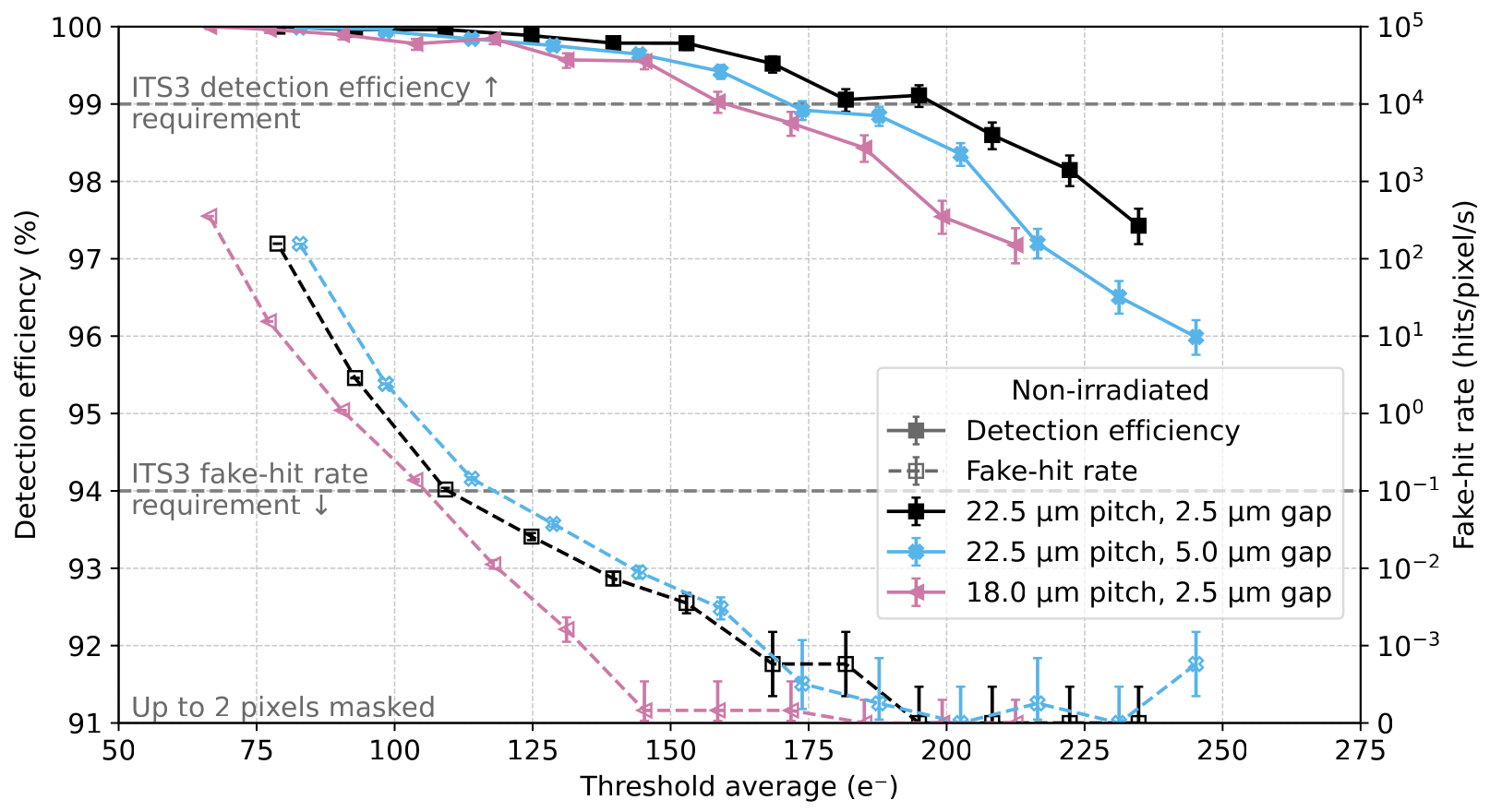}
    \caption{Performance of the irradiated and non-irradiated MOSS sensor measured during a test beam at CERN PS \cite{MOSS_preprint}.}
    \label{fig:MOSS_efficiency}
\end{figure}
\noindent The MOSS is a 6.7-megapixel sensor with dimensions of $1.4 \times 25.9 \text{ cm}^2$. It consists of ten Repeated Sensor Units (RSUs), each comprising two independently powered half-units (top and bottom) featuring different pixel architectures. 
These components are patterned on the same reticle and exposed sequentially onto the wafer, ensuring electrical continuity across boundaries. This technique eliminates the need for a flexible circuit on top of the sensor.
The MOSS was specifically designed to evaluate the manufacturing yield and demonstrate the feasibility of power and signal transmission across a length of \SI{26}{cm}.\\
Data can be read out either locally from an individual half-unit or from the left endcap via the stitched backbone. The sensor underwent comprehensive characterization both in the laboratory and under test beam conditions to evaluate its detection efficiency, as well as its spatial, timing, and energy resolution - before and after irradiation (Total Ionizing Dose [TID] and Non-Ionizing Energy Loss [NIEL]). Figure \ref{fig:MOSS_efficiency} demonstrates that MOSS meets its design goals, achieving a detection efficiency exceeding 99\% and a fake hit rate below \SI{e-6}{hits/pixel/event}, even after exposure to ALICE-relevant radiation levels of \SI{4}{kGy} (TID) and \SI{4e12}{1\,MeV n_{eq}/cm^2} (NIEL) \cite{MOSS_preprint}.

\begin{wrapfigure}{r}{0.63\textwidth}
    \centering
    \includegraphics[width=0.6\textwidth]{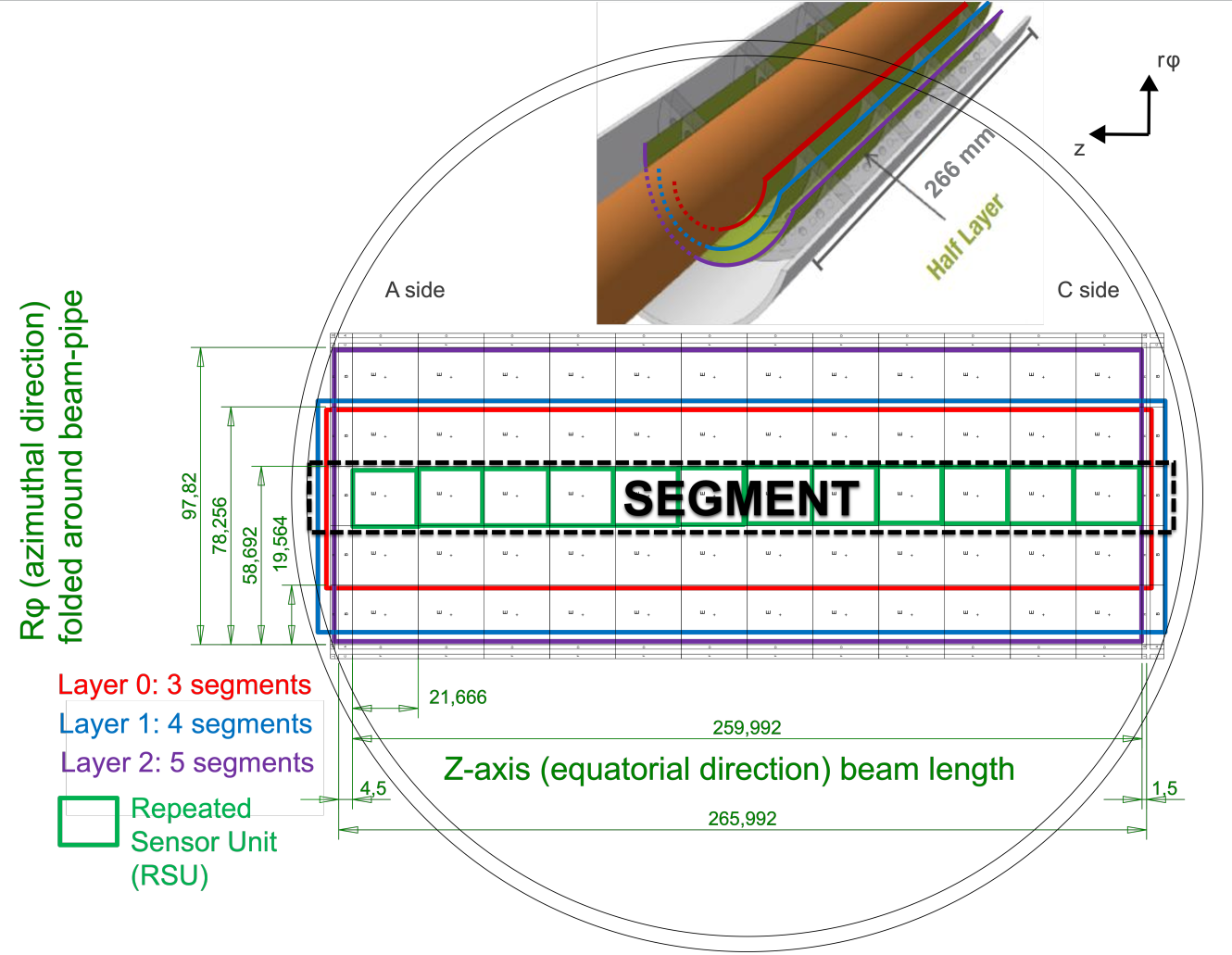}
    \caption{Floor-plan of the final sensor production \cite{ITS3_TDR}.}
    \label{fig:Stitching_MOSS}
\end{wrapfigure}

The final ITS3 sensor will contain 3, 4, or 5 independent detector segments — called MOnolithic Stitched Active pIXel (MOSAIX) — measuring $19 \times 266 \text{ mm}^2$.
As shown in figure \ref{fig:Stitching_MOSS}, each segment consists of a left endcap with the readout and power interfaces, 12 Repeated Sensor Units (RSUs) with pixel matrices, and a right endcap with additional power connections.
Each RSU is further divided into 12 pixel sub-matrices ($20.8 \times 22.8$ \SI{}{\micro m^2} pitch), which can be powered and operated independently. The wafer is thinned to \SI{\sim 50}{\micro m}, making the silicon flexible enough for a self-supporting curved geometry. The three layers per half-barrel are then formed by thinning, dicing, and bending the sensors.
\section{Mechanical and thermal design}
\label{sec:mechanics_cooling}
\begin{wrapfigure}{r}{0.7\textwidth}
    \centering
    \includegraphics[width=0.65\textwidth]{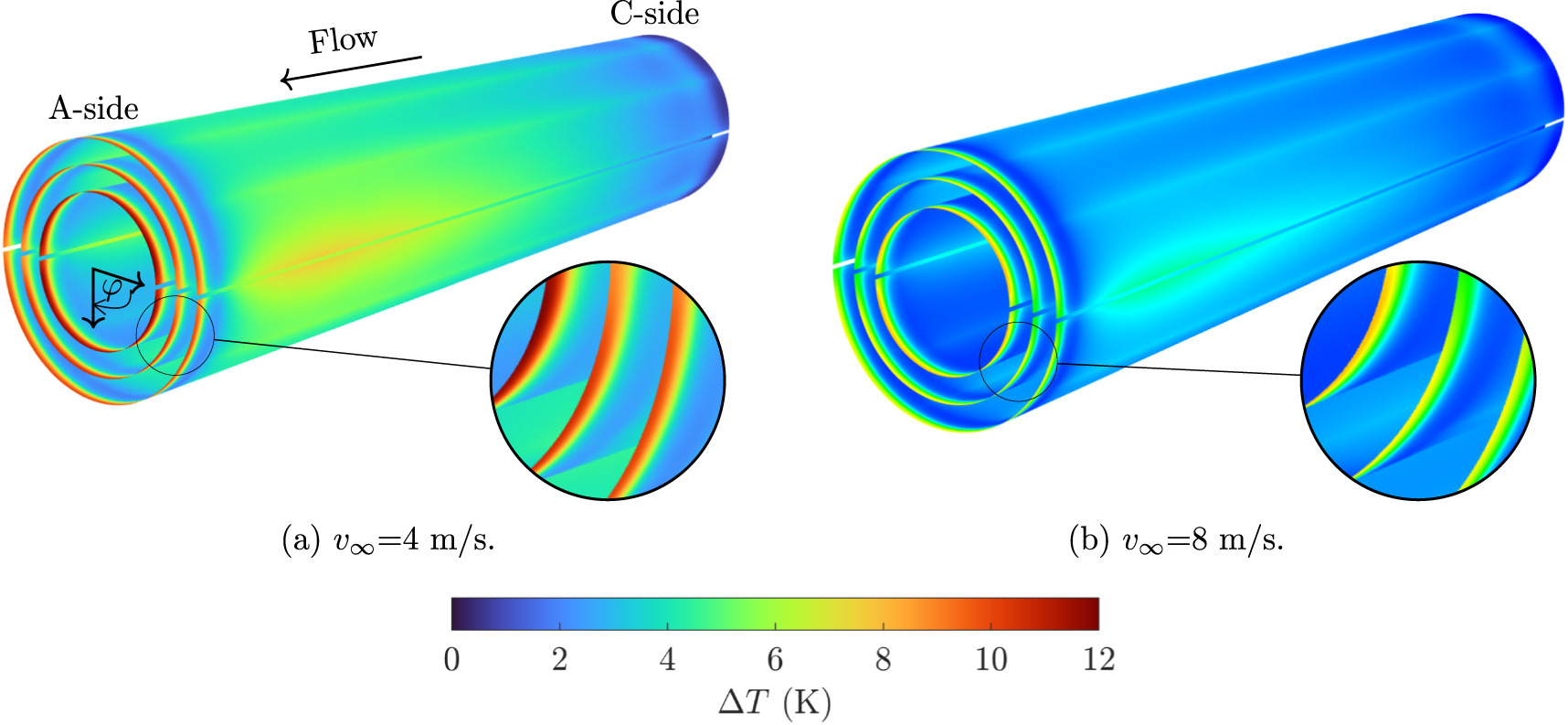}
    \caption{Temperature variation measured along the sensor for the three ITS3 layers \cite{ThermalTest_ITS3}.}
    \label{fig:Temp_ITS3}
\end{wrapfigure}
After demonstrating the possibility to bend a silicon sensor, the question of whether it was possible to assemble a tracker cooled by air only while granting mechanical stability arose.
Moreover, silicon sensors designed for the ITS3 are expected to have a non-uniform heat dissipation since the readout electronics will be placed on one side of the sensor. 
To address this last issue, a more conductive carbon foam (K9) was used on the A-side of the sensor. Since the semi-cylindrical support just on the A-side of the sensor was not sufficient for mechanical stability, other carbon foams (RVC) with less thermal conductivity were added on the edges of the sensors (longerons and C-side rings) \cite{ThermalTest_ITS3}. Several simulations and experimental measurements were conducted to test the thermal uniformity of the sensor. A setup was built to simulate the expected heat dissipation and thermal conductivity of the future ITS3. Different measurements were taken with different air fluxes: as shown in figure \ref{fig:Temp_ITS3}, an air flux of \SI{8}{m/s} is required to have a temperature gradient of less than \SI{10}{K} along the detector. Further details can be found in \cite{ThermalTest_ITS3}.
\begin{figure}[!ht]
    \centering
    \includegraphics[width=0.9\textwidth]{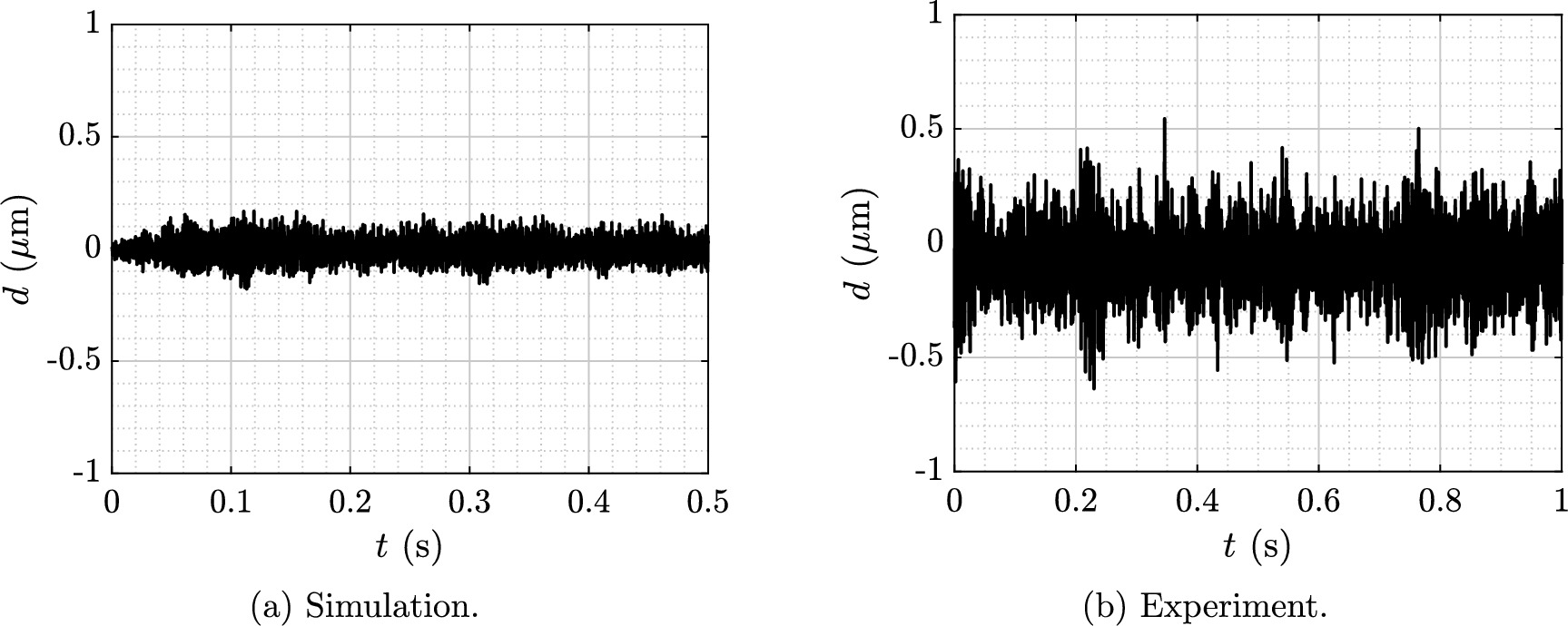}
    \caption{Displacement of the outer layer measured at the center of the sensor \cite{VibrationalTest_ITS3}.}
    \label{fig:Vib_ITS3}
\end{figure}
The next issue addressed is whether an air flow of \SI{8}{m/s} gives a displacement negligible with respect to the sensor spatial resolution. To prove this point, a mock-up version of the ITS3 was built and tested for vibrations produced by air cooling. The results, reported in figure \ref{fig:Vib_ITS3}, show a displacement of \SI{1}{\micro m} peak to peak.
Simulation results are more reliable than the experimental ones since experimental data are affected by a low frequency vibration related to the fan that generates the airflow. The prototype also differs from the final detector, but with simulation it was possible to tune the foam design to find the required balance between structural displacements, energy consumption, and detector
cooling \cite{VibrationalTest_ITS3}.
\section{Conclusions}
ALICE will deploy the ITS3 detector for LHC Run 4, marking the development of the first truly cylindrical wafer-scale Monolithic Active Pixel Sensor. It has been demonstrated that silicon can be bent without compromising detector performance. Both thermal uniformity and mechanical stability have been validated through simulations and experimental tests. Sensor prototypes achieved detection efficiencies exceeding 99\% at the ALICE ITS3 target fluence of \SI{4e12}{1\,MeV n_{eq}/cm^2} and \SI{4}{kGy}, even when operated at room temperature without active cooling.
The first stitched sensors have been successfully tested and the final prototype is currently in production.
\bibliographystyle{JHEP}
\bibliography{biblio.bib}
\end{document}